# Big Variates – Visualizing and identifying key variables in a multivariate world

S. J. Watts, L. Crow

School of Physics and Astronomy, University of Manchester, Oxford Road, Manchester, M13 9PL, UK

**ABSTRACT**

Big Data involves both a large number of events but also many variables. This paper will concentrate on the challenge presented by the large number of variables in a Big Dataset. It will start with a brief review of exploratory data visualisation for large dimensional datasets and the use of parallel coordinates. This motivates the use of information theoretic ideas to understand multivariate data. Two key information-theoretic statistics (Similarity Index and Class Distance Indicator) will be described which are used to identify the key variables and then guide the user in a subsequent machine learning analysis. Key to the approach is a novel algorithm to histogram data which quantifies the information content of the data. The Class Distance Indicator also sets a limit on the classification performance of machine learning algorithms for the specific dataset.

1. Introduction

The analysis of large data sets is big business and is fundamental to extracting science from experiments in astronomy, life sciences, and particle physics. So called "Big Data" is now a key commodity in science, healthcare, business, and industry. Most datasets can be written as a spreadsheet with variables (P columns) and instances/events (N rows). This paper will concentrate on the analysis of data with a large number of variables, described in the title as *Big Variates*. The key idea is to visualise multivariate data and the inter-relationships between the variables in a model independent manner, which can be used to guide further analysis using modern data mining algorithms. Central to the methodology is the consistent use of Claude Shannon's information theory. The data is treated as an information source, and information-theoretic statistics are derived to quantify the relationships between the variables. This paper will show how to perform an *"Information theoretic multivariate analysis"*.

This paper is organised into a brief review of multivariate data visualisation, the methodology to calculate two information theoretic statistics after binning the data using a novel histogram algorithm, and then a guided analysis using these statistics to identify the key variables in the dataset. This analysis is then compared with a conventional machine learning approach.

2. Visualising Multivariate Data

Histograms (1D) and scatter plots (2D) are the most widely used methods when undertaking an Exploratory Data Analysis (EDA). Getting a "feel" for the data is normally the first task of data analysis. This is especially important to check that data has been recorded correctly and to investigate outliers. It also initiates the process of deciding what statistical techniques ( e.g. curve fitting) might then be followed. One can extend these plots to 3D by using perspective. Colour and icons can also be used to extend the information provided to the viewer. Scatter plot matrices provide a useful tool, but the number of these grow rapidly with the number of variates, P(P-1)/2. To visualize data with dimensionality greater than three is very difficult. There are two geometric methods that preserve the original dimensionality of the data (number of variates), and allow individual points to be identified – parallel coordinates [1,2] and polyviz [3]. An excellent review discusses this issue in more detail [3]. This paper concentrates on parallel coordinates. The technique was first invented in 1885 and re-invented a century later. The advent of modern computers, graphics cards and displays makes the technique viable. It produces more than a picture since its mathematical structure can be exploited to solve problems [4].

2.1 Parallel Coordinates

A "parallel coordinates" plot visualizes points in a space of dimension P by mapping them onto a two dimensional plane in which the axes are placed parallel to one another. Essentially the P-dimensional point is mapped to a line segment on the 2D plane. Figure 2.1 illustrates five points in a 4D space. These points are given in the table included in the figure. One is highlighted using a dashed line. This is called "brushing" and colour can be used to highlight interesting points. The ability to interact with this diagram, via a computer generated Graphical User Interface (GUI) involving brushing and removing selected points ("pruning") makes this a powerful way to understand the data from a single image. The GUI has to be well designed for this to work well. Reference [5] describes the use of parallel coordinates in data analysis.

2.2 Application to a Particle Physics Monte Carlo (PPMC) dataset

A Monte Carlo dataset for the decay of a $K_s^o$ -> $\pi^+\pi^-$ was supplied by the author of reference [6]. Further details about this dataset can be found in this reference. There are 1264 and 3736 signal and background events respectively in the sample. Each event has eight variables plus one class variable:

- Doca – distance of closest approach between the two pions
- $R_{XY}$ – half length of the cylinder defining the $e^+e^-$ interaction region from which $K_s^o$ emerges
- $|R_z|$ – half length of the cylinder interaction region in the z direction.
- $|Cos(\Theta_{hel})|$ – absolute value of the cosine of the $K_s^o$ helicity angle
- Sfl - $K_s^o$ signed flight length
- Fsig – statistical significance of the $K_s^o$ flight length
- Pchi – chi-squared probability of the $K_s^o$ vertex
- Mass – $K_s^o$ reconstructed mass
- Flag – class variable. 0 if background, 1 if signal.

The analysis in this paper does not make any physics assumptions.

The parallel coordinates plot is shown in Fig. 2.2. Data visualisation benefits hugely from the use of colour, as this figure illustrates. The signal and background events have been brushed red and green respectively. Colours are defined with three (red, green and blue ) coordinates. In addition, graphics cards have an "alpha channel", which defines the degree of transparency of each colour. This can be used to blend colours. In Fig. 2.2, areas of the plot in which signal and background events merge are a blend of red and green, which gives yellow. The plot is also more intense when the density of points is high. The intensity can be controlled by altering the value of the alpha channel parameter.

In a single plot, one can quickly understand key features of this eight variable dataset.

- The only variable in which the signal shows some separation from background is Fsig.
- For |Cos($\Theta_{hel}$)| and Pchi, the signal and background events merge.
- For Doca, Rxy and |$R_z$| and Sfl, the background events are discriminated from signal in certain regions.
- The mass of the signal events appears in a clear range, but is contaminated by background. The overlap region is yellow.

This plot motivates the key question this paper addresses – *what are the key variables ?* A visual Exploratory Data Analysis is useful because the human visual system is so powerful. However, it is a potential source of systematic bias. This paper proposes computer programmed algorithms using the data as input, to identify the key variables. This is a data driven analysis which is model independent.

## 3. Information Theoretic Measures

To answer the question posed above one needs to find statistics that can be estimated from the data that apply no matter how many variables are involved. The parallel coordinates plot is similar to Claude Shannon's diagrams in Figure 12 of reference [7]. These show the probability with which source signal values are transmitted to recorded values at the receiver. This suggests using ideas from information theory. Information is measured in Shannon's theory using the Shannon Entropy, defined in Eq. 1.

$$H(bits) = -\sum_{i=1}^{i=K} p_i \, log_2 p_i \tag{1}$$

where $p_i$ are the probabilities for each of the K possible values. This is the discrete entropy and it has several key properties; it is always greater than zero, is additive for independent events, and can be calculated for any variable either individually or in combinations. For discrete variables - so-called categorical data – it is trivial to calculate H. For continuous variables, one must first histogram the data. One can define the differential entropy, h, for a continuous probability density function (pdf), p(x), as follows,

$$h \, (bits) = \frac{1}{log_e 2} \int_{-\infty}^{+\infty} -p(x) log_e p(x) dx \tag{2}$$

h does not have the same useful properties as H. It measures the "spread" of the data. For example, h is proportional to log($\sigma$) for a Gaussian distribution, where $\sigma$ is the standard deviation. It generalises the concept of the standard deviation to any number of dimensions.

Two key information theoretic statistics that can be calculated from the data are described in Sections 3.1 and 3.2. One measures the shared or mutual information between variables. The other measures how different the probability distributions are for two classes of event.

3.1 Similarity Index

Figure 3.1 shows schematically the sharing of information between two variables X and Y. This is called the Mutual Information, I(X,Y). Ref. [8] provides a more detailed description. The entropies of variable X, variable Y, and the combined variables X and Y, are H(X), H(Y) and H(X,Y) respectively. The probability distributions of the variables required for the calculation are p(X), p(Y) and p(X,Y) respectively. If H(X,Y) = H(X) + H(Y), then the variables are independent. If H(X,Y) is the Min(H(X),H(Y)), then the variables are completely dependent. Thus one can define a "Similarity Index" which measures the fraction of shared information – see Figure 3.1. This is the information theoretic version of the correlation coefficient and has a value between zero and one. It can only be estimated from binned data. The formula for the Similarity Index (SI) is,

$$SI = \frac{I(X,Y)}{Min(H(X), H(Y))} \qquad (3)$$

When identifying a key variable, one needs it to share information with other variables. Variables that do not share any information are not useful to understanding the data and can be discarded.

3.2 Class Distance Indicator

If one separates the samples/events in the dataset into two different classes, it is important to know how different the probability distribution of events of one class, labelled 1, are from events from another class, labelled 2. A well-known measure of this is called the Kullback-Leibler Divergence or Kullback-Leibler Distance or Relative Entropy [9], which has units of bits, if the logarithm is to base 2.

The Kullback-Leibler distance is defined as,

$$KL(p_1, p_2) \equiv \int p_1(x) log\left(\frac{p_1(x)}{p_2(x)}\right) dx \qquad (4)$$

where, $p_1(x)$ and $p_2(x)$ are the pdf's for Class 1 and Class 2 events respectively. For simplicity only the single variable version is defined, but one can calculate this for a multi-variable space.

In fact there are two "distances", $KL(p_1, p_2)$ and $KL(p_2, p_1)$ which may not be the same. This is because the distance is normalised to the "spread" of Class 1 or 2 events. In other words, it is a normalised distance, which is why this paper will refer to the non-parametric estimate of the Kullback-Leibler distance as the "Class Distance Indicator (CDI)." Using an analogy, the Sun-Earth distance is 149.6 million km. The diameter of the Sun and Earth are 1.39 million km and 12,700 km respectively. The "distance indicator" between the Sun and Earth is 107 Sun diameters or 11780 Earth diameters. It is easier to detect the Sun than the Earth when looking from outside the Solar System.

A non-parametric estimate for the Kullback-Leibler distance in Eq. (4), CDI(1,2) is, [10]

$$CDI(1,2) = \frac{P}{N_1}\sum_{i=1}^{N_1} log_2\left(\frac{\lambda_i^{12}}{\lambda_i^1}\right) + log_2\left(\frac{N_2}{N_1-1}\right) \quad \text{bits} \quad (5)$$

where, $N_1$ and $N_2$ are the number of events in Class 1 and Class 2 respectively
$\lambda_i^1$ is the nearest neighbour distance between the i[th] point in Class 1 and all other Class 1 points.
$\lambda_i^{12}$ is the nearest neighbour distance between the i[th] point in Class 1 and all Class 2 points.

Note that this estimate can be made for any number of variables in the data space. The second log term adjusts for imbalanced classes, i.e. $N_1 \neq N_2$. CDI can be calculated for either continuous or binned data as Equ. (4) is scale invariant.

The non-parametric estimate calculates the average log of the nearest neighbour distances between the points. When applied to a single class of events, this is a measure of the spread (variance) of the points, and consequently the differential entropy – see also Section 3.3. When applied to two different classes of events, it measures how far apart they are. If the distributions are the same, then one obtains the same value and thus the CDI, which is the log of the ratio of these two measures, becomes zero – and then one cannot distinguish the two classes of events.

The KL divergence and its estimate, CDI, place a fundamental limit on how well one can separate two classes of events. This is called Stein's Lemma – although due to Chernoff, cf. reference [11]. For a large number of events, involving two classes, Signal (S) and Background (B),

$$Probability\ of\ false\ alarm \approx 2^{-CDI(S,B)} \quad (6)$$

If the underlying signal and background distributions are identical, CDI(S,B) = CDI(B,S) = zero, and the probability is 100%. No matter how clever the machine learning algorithm, one cannot beat the limit given by Equ. (6).

3.3 Histogram Algorithm

Variables are either categorical/discrete, or continuous. Continuous variables must first be made discrete by generating a histogram of the data. One can then calculate the Similarity Index introduced in Section 3.1. It is critical for consistency that the same histogram algorithm is applied to all continuous data. The histogram algorithm used in the analysis is described briefly in this Section. The relation between the discrete (H) and differential (h) entropies defined in Eq. (1) and Eq. (2) respectively is [9],

$$H = h - log_2 \Delta \quad (7)$$

where $\Delta$ is the bin width. h is well defined because it depends on the probability distribution. H appears to depend on the chosen bin width. However, the bin width is not arbitrary. It must be chosen to ensure that the histogram is not over-binned or under-binned. Over-binning occurs when the bin size is too small, and Poisson fluctuations affect the quality of the histogram. Under-binning occurs when the bin width is too large, and the probability density function is poorly estimated. The optimal bin width can be found by demanding a minimum integrated squared error (MISE) between the underlying distribution and the one estimated by the histogram. The classic solution is due to Scott, ref. [12]. However, one first needs to know the underlying distribution, and the Scott solution fails if the first derivative of the pdf is zero, which applies in the case of a uniform distribution. It is simple to show that for many well-known distributions ( Gaussian, exponential, triangular, Maxwell-Boltzmann), using Scott's equation for the bin width, and calculating h, that H is (1/3)log($\beta$N), with $\beta$ close to one.

An ansatz is thus made to set,

$$H = \frac{1}{M} log_2 N \qquad (8)$$

with M > 1. For a uniform distribution one can show exactly that provided N > 35, Poisson fluctuations are removed for M > 2. Monte Carlo simulations with different distributions indicate that M must be in the range 2 to 3 – discussed further below. Since M = 2 gives the largest value of H - most information – and has no Poisson fluctuations, this provides a well-defined H for each variable.

Now that H is well-defined, by estimating h, one can determine the bin size, $\Delta$ ,

$$\Delta = \frac{2^{h(bits)}}{N^{\frac{1}{M}}} \qquad \text{which for M = 2 is} \qquad \Delta = \frac{2^{h(bits)}}{\sqrt{N}} \qquad (9)$$

h is estimated with the non-parametric estimator of Kozachenko and Leonenko, [13], which uses the nearest neighbour distance, $\lambda_i$, for each point,

$$h = \frac{1}{N}\sum_{i=1}^{N} log_2 \lambda_i + log_2[2(N-1)] + \frac{\gamma}{log_e 2} \qquad (10)$$

where $\gamma$ = 0.5772, is the Euler–Mascheroni constant.

Equ. (9) and Equ. (10) are simple to use, and make no assumption about the underlying pdf. For a uniform distribution, Equ. (9) gives,

$$\Delta = \frac{Range}{\sqrt{N}} \qquad (11)$$

Reference [14] defines a cost function as a function of bin size. This penalises over-binning and under-binning. Its minimum provides the optimal bin size. The cost function is,

$$Cost = \frac{2\mu_B - \sigma_B^2}{\Delta^2} \qquad (12)$$

Where, μ$_B$ and σ$_B$ are the mean number of events per bin and associated standard deviation respectively. The cost function is calculated as M is varied between 1 and 6 for uniform, Gaussian and exponential distributions, using 5000 Monte Carlo generated events, and is shown in Fig. 3.3. The cost function has been shifted such that its value at M = 2 is zero, and the overall range is scaled to one for the cost function between M = 1 and M = 2. Fig 3.3 shows that the cost function drops rapidly between M = 1 and M = 2 as Poisson fluctuations are removed. The cost function for the uniform distribution does not increase at larger M – it is a flat distribution so there is no cost increase for larger bin size, but it does show fluctuations as binning digitises the data. These fluctuations are negative and are thus not seen on the log scale. Fig 3.3 shows clearly that all distributions behave in the same way using a scaled and shifted cost function. The over-binning between M = 1 and M = 2 is clear. Beyond M = 3, the cost function increases due to under-binning.

Figure 3.4 shows how the algorithm performs on Gaussian data for M = 1, 2, 3, and 6. This shows how M identifies histograms with over-binning and under-binning. M of between 2 and 3 provides a robust estimate of the underlying probability distribution function. Equ. (8), Fig. 3.3 and Fig. 3.4 illustrate the comment in ref. [15] of Eadie et al., " too few bins carry little information, but to many bins lead to too few events per bin ".

3.4 Information content of data

Section 3.3 shows that for continuous data, the maximum information content is $\frac{1}{2}log_2 N$ bits per variate. If the number of events is increased by a factor 4, then the information increase is *"one bit per variate"*. This is the information theoretic version of *" four times more data reduces the error by a factor 2, provided one is statistics limited "*. For example, if there are 1024 events and one variable, then the information content is 5 bits. This increases to 6 bits with 4096 events. However, if another variable can be identified, then for 1024 events the information content will increase to 10 bits, minus any shared information. In conclusion, in some cases, it is better to find more relevant variables than record more data. Moreover, if the data has two classes ( e.g. Signal and Background ) even if the Class Distance Indicator is theoretically X bits – based on a prior knowledge of the underlying distributions - one will not achieve the separation of Signal and Background limit in Equ. (6) until one has recorded data with X bits worth of total information.

4. Application to the PPMC dataset

The particle physics dataset described in Section 2.2 is analysed using concepts from Section 3. Each variable is made discrete by histogramming using Equ. (9) and Equ. (10) so that the Similarity Index, Section 3.1, can be calculated. The Class Distance Indicator is also calculated using Equ. (5) for all possible combinations of the variables. The results of these calculations are described below.

### 4.1 Variable Interaction Diagram

The results of the Similarity Index calculation are displayed in Fig. 4.1 such that the interactions between the variables can be clearly identified. This we call a *"Variable Interaction Diagram"*. Each variable is marked with a labelled dot on the circle. Variables with significant shared information are linked by a line. The class variable is placed in the centre of the circle. Colour can be used to indicate the value of the Similarity Index, however in this black and white schematic, links with SI > 0.25 and 0.04 < SI < 0.1 are shown by filled and dashed lines respectively.

There are strong links between the class variable for Fsig, Sfl and Mass, and also between Fsig and Sfl. Four weaker links are found between Doca and PChi, Rxy and Sfl, Doca and Sfl, and Rxy and Fsig. These values are given in Table 4.1 along with the CDI estimate.

### 4.2 Class Distance Indicator

Table 4.1 shows the CDI for all variables. The CDI has been ranked and only top-ranked values are shown for variable combinations above two. The % of correctly classified events using machine learning algorithms is also tabulated. Section 4.3 describes this further. In addition, a quantity called CDR is shown. This is the parallel combination of CDI(B,S) and CDI(S,B), for which the equation is,

$$\frac{1}{CDR} = \frac{1}{CDI(S,B)} + \frac{1}{CDI(B,S)} \tag{13}$$

In a two class classification problem the machine learning algorithm normally does its best to optimise the performance for both classes. In this case, a better estimate of the CDI to use in Equ. (6) is the CDR defined in Equ. 13. See reference [11] for more details.

Fig. 4.1 and Table 4.1 summarise the key information theoretic statistics to understand this data and to identify what variables are most effective in machine learning. Fig 4.1 and Table 4.1 immediately lead to the following conclusions.

- The SI between each variable and the class picks out key variables to discriminate signal and background – Fsig, Sfl and Mass. This is confirmed by the CDI values which are significant for these variables.
- The combination of the SI and CDI values guides one to key plots. This is illustrated in Fig 4.2 which shows scatter plots for various variables. Fig 4.2a shows Pchi versus Rz. The SI is zero and thus there is no relationship between the variables, but there is a small value of CDI which is slightly larger for CDI(B,S) at 0.4 bits. The plot shows that some background – but not all – can be separated from signal at certain values of Rz. Fig 4.2b shows Cos-Hel versus PChi. Both SI and CDI are zero. There is nothing of interest at all in this plot. Fig 4.2c shows Sfl versus Fsig. SI is significant, and so are the CDI values at ~ 4 bits. The plot shows that the data has structure in these variables, and that signal and background can be discriminated.

**Table 4.1 Summary of SI and CDI values for the Particle Physics Monte Carlo Dataset.**

| Variables | # | SI Note 1 | CDI(B,S) Bits | CDI(S,B) Bits | CDR Bits | Correctly Classified (%) |
|---|---|---|---|---|---|---|
| Fsig | 1 | 0.46 | 4.03 | 3.13 | 1.76 | 90.22 |
| Sfl | 1 | 0.27 | 3.66 | 2.71 | 1.56 | 88.88 |
| Mass | 1 | 0.28 | 2.0 | 1.58 | 0.88 | 80.98 |
| Doca | 1 | - | 0.28 | 0.26 | 0.14 | 72.84 |
| Rxy | 1 | - | 0.29 | 0.14 | 0.10 | 71.14 |
| Rz | 1 | - | 0.18 | 0.12 | 0.07 | 71.3 |
| Pchi | 1 | - | 0.11 | 0.12 | 0.05 | 70.92 |
| Cos-Hel | 1 | - | 0.05 | 0.03 | 0.02 | 70.8 |
| Rxy/Fsig | 2 | 0.04 | 4.46 | 5.10 | 2.38 | 95.18 |
| Doca/Fsig | 2 | 0.06 | 4.25 | 4.67 | 2.22 | 93.44 |
| Sfl/Fsig | 2 | 0.25 | 4.45 | 4.17 | 2.15 | 93.92 |
| Rxy/Sfl | 2 | 0.06 | 4.20 | 4.32 | 2.13 | 93.58 |
| Rxy/Sfl/Fsig | 3 | NA | 4.93 | 6.02 | 2.71 | 95.26 |
| Doca/Rxy/Fsig | 3 | NA | 4.75 | 6.04 | 2.66 | 95.3 |
| Rxy/Cos-Hel/Sfl/Fsig | 4 | NA | 5.54 | 6.79 | 3.05 | 95.24 |
| Doca/Rxy/Sfl/Fsig | 4 | NA | 5.38 | 6.98 | 3.04 | 95.34 |
| Doca/Rxy/Cos-Hel/Sfl/Fsig | 5 | NA | 5.94 | 7.55 | 3.32 | 95.40 |
| All except Rz and Mass | 6 | NA | 6.48 | 7.63 | 3.50 | 95.36 |
| All except Mass | 7 | NA | 6.45 | 7.79 | 3.53 | 95.34 |
| All | 8 | NA | 7.09 | 8.93 | 3.95 | 95.74 |

Note 1: The SI values for one variable use Class as the second variable. In the SI column, "– " means zero and "NA" means it is not defined for three or more variables.

- The next most interesting relationships are those involving two variables. The most important of which is Rxy versus Fsig. The SI is small but non-zero and there is a significant CDI value which is larger than the CDI for the individual variables. Rxy shows no relationship to the class variable alone. However, when paired with Fsig there is a significant relationship. This is because there is a significant three-way interaction between Rxy, Fsig and the Class variable. One can measure this by using an information theoretic statistic called the "*Interaction Information (II)* ". For two variables this is the Mutual Information, which in this paper has been normalised and is called the Similarity Index. For three variables, we will use the Interaction Information defined by ref. [16] as co-information, which is the same as that used by ref. [17] when defining higher order mutual information for studying interactions in spin systems. Ref. [18] has a full discussion on this subject and is one of the first papers to use information theory considerations when quantifying attribute interactions. The 3-way Interaction Information, [16, 17], is thus,

$$II(A,B,C) = H(A) + H(B) + H(C) - H(A,B) - H(A,C) - H(B,C) + H(A,B,C) \quad (14)$$

For Rxy, FSig and Class, II is around 0.5 bits. In other words, the relationship between Rxy and Class is only revealed when combined with Fsig. This is shown in Fig 4.2d which clearly shows that a particular region of the Rxy/Fsig space separates signal from background.

- To conclude, in this supervised learning problem, in order to select the class and separate signal from background – one first selects variables directly linked to the class ( Sfl and Fsig) and then those connected to these variables ( Rxy and Doca). The Mass variable is not used, as this variable measures the $K_s^o$ reconstructed mass, and is not included to avoid bias. Table 4.1 clearly shows that the key discriminating variables are Rxy, Sfl and Fsig. Adding in variables beyond this point has a marginal effect on the CDR which flattens out at ~ 3 bits.

Finally, note that there are $2^P$ possible combinations of CDI to estimate. The Variable Interaction Diagram requires P(P-1)/2 Similarity Index calculations. This diagram substantially reduces the number of combinations one needs to consider to find the relevant variables to discriminate signal from background.

### 4.3 Comparison with datamining algorithms

The WEKA data mining suite, Ref [19], was used to analyse the PPMC dataset. The 1R algorithm was used for single variables and the J48 decision tree algorithm for multiple variables. Both algorithms are described in ref. [19]. 1R is an algorithm that classifies events on the basis of a single attribute, i.e it is a 1-level decision tree. J48 is an open source Java implementation of the C4.5 algorithm in the WEKA data mining suite which uses information gain when building a decision tree. Other algorithms were tried for multiple variables (e.g. neural net, support vector machine), however these could not match the effectiveness of the decision tree algorithm for this specific dataset. Table 4.1 shows the percentage of correctly classified events of both type,

$$Percentage\,(\%) = \frac{True\,Signal + True\,Background}{All\,Events} \qquad (15)$$

Even with a random choice, this cannot fall below 70% because the signal-to-background ratio is around 1/3. The classification success does not go above 95%. The datamining also agrees on the relevant variables. The false-positive rates for signal and background are ~ 3% and ~ 9% respectively. This implies a CDI of around 5 bits to 3.5 bits, which is consistent with Table 4.1. The information-theoretic analysis shows that there is no more that the data mining software can do. There is not enough information in the data to get better signal to noise selection.

### 5. Conclusions

This paper has proposed a new methodology for data analysis. Starting with a new dataset, one first makes an exploratory visual analysis using a parallel coordinates plot. After applying a new binning algorithm that ensures that each variable has a fixed information content, one then calculates the mutual information between variables. This leads to a Variable Interaction Diagram which is used to identify the key variables in the dataset. In a supervised learning problem, this then identifies which variables to use that will maximise the Class Distance Indicator and thus the separation between classes of events. The CDI also sets a limit on the classification performance of machine learning algorithms for the specific dataset. The analysis is data driven and is model independent. It can also be used to guide the user to the relevant histograms and scatter plots. For data sets containing a large number of variables, this type of analysis should significantly reduce analysis time and guide a user to the correct conclusions about the nature of the data.

**Acknowledgments**

We thank Dr L. Teodorescu from Brunel University for providing the data set analysed in this paper. We thank Benjamin Radburn-Smith, Stuart Phillips and Zachary Baker for their work on earlier versions of the DataViewer software. Support from the UK Research and Innovation, Science and Technology Facilities Council (STFC) is gratefully acknowledged.

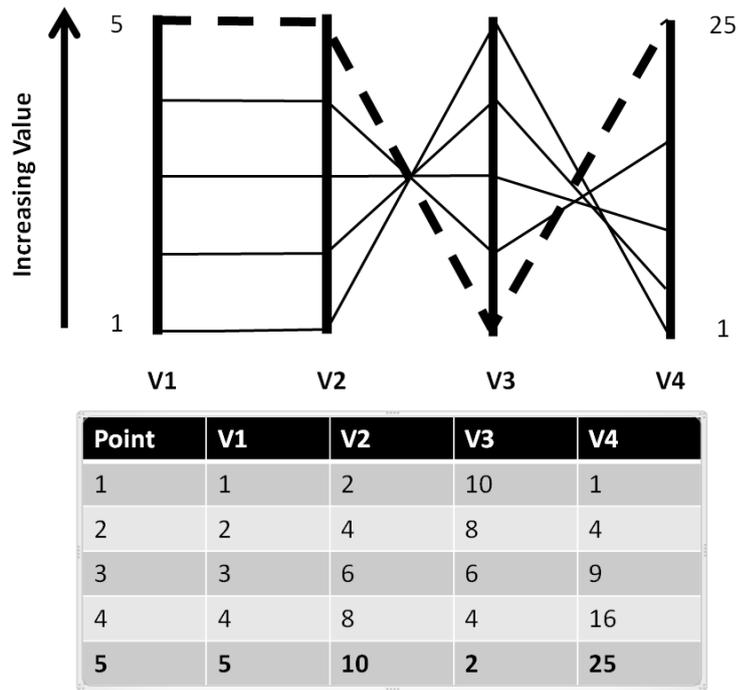

| Point | V1 | V2 | V3 | V4 |
|-------|----|----|----|----|
| 1 | 1 | 2 | 10 | 1 |
| 2 | 2 | 4 | 8 | 4 |
| 3 | 3 | 6 | 6 | 9 |
| 4 | 4 | 8 | 4 | 16 |
| 5 | 5 | 10 | 2 | 25 |

Fig 2.1   Diagram to illustrate a parallel coordinates plot. There are four variables (V1,V2,V3 and V4) and five points. The table gives the point coordinates which are shown in the plot above. The fifth point has been highlighted by using a dashed line segment – this is called "brushing".

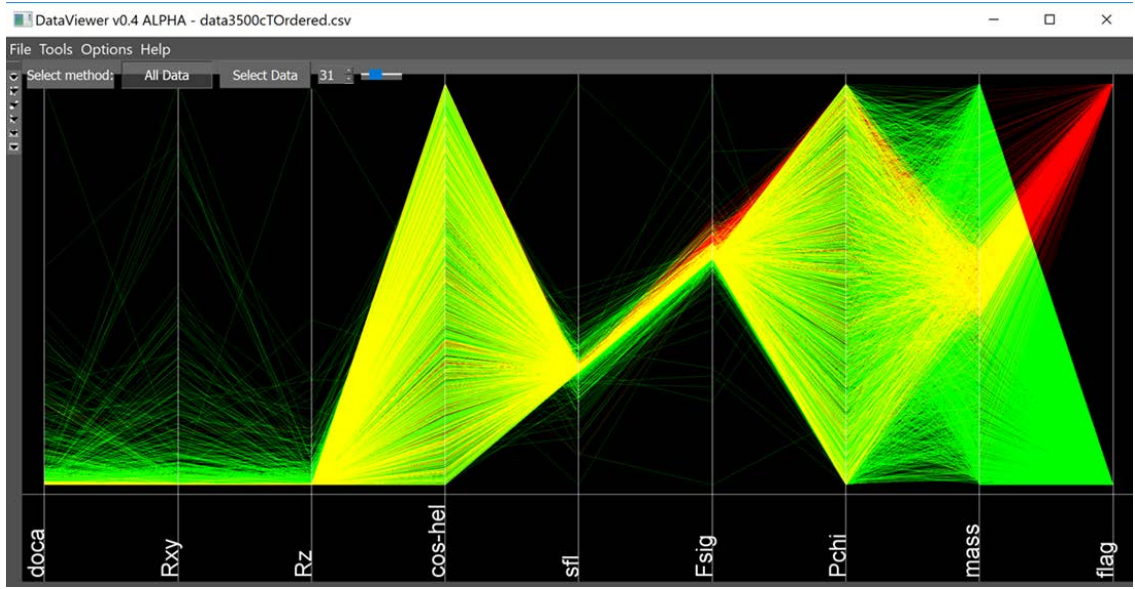

Fig 2.2  Parallel coordinates plot of the particle physics Monte Carlo data described in the text using the DataViewer software. Signal and background points are brushed RED and GREEN respectively. The variables are given at the bottom of the plot and are described in the text.

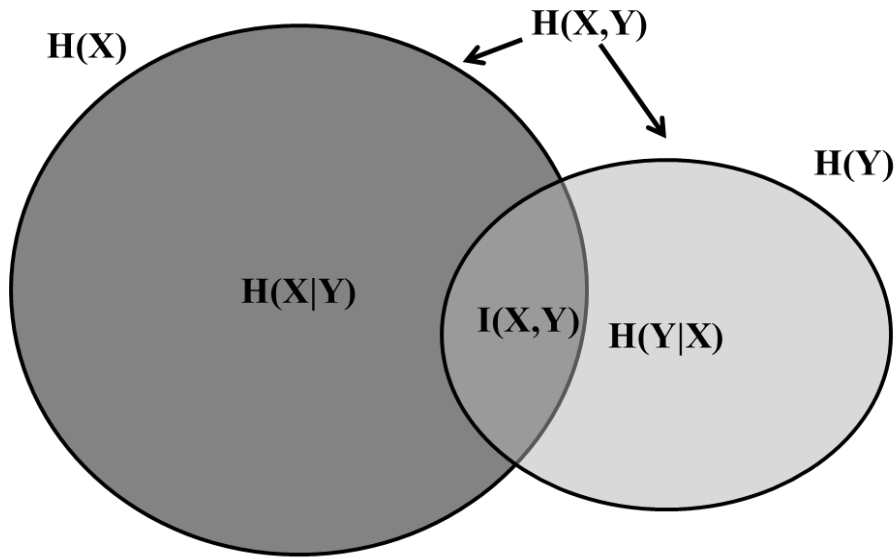

$I(X,Y) = I(Y,X) = H(X) + H(Y) - H(X,Y)$   Mutual Information

Similarity Index, $SI = I(X,Y)/(Min(H(X), H(Y)))$

SI = 1   Variables are completely dependent
SI = 0   Variables are completely independent

Fig 3.1 Venn diagram explanation of the relationship between the entropy (H) for variables X and Y and the relationship between H(X), H(Y), H(X,Y), H(X|Y), H(Y|X), and the mutual information I(X,Y).

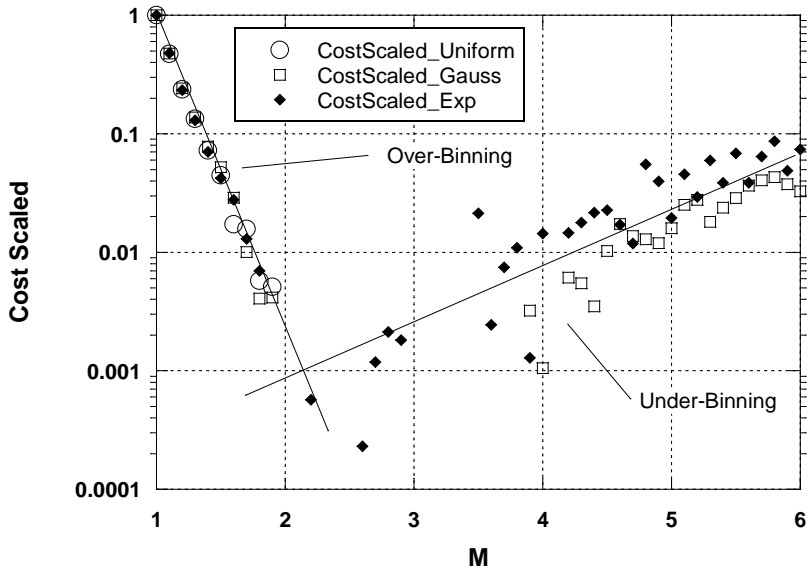

Fig 3.3 Cost function versus M for a Uniform distribution (range from zero to one), Gaussian distribution ( mean zero, standard deviation of one), and Exponential distribution ( mean = standard deviation = one).  The legends are Open Circle, Open Square, and Filled Diamond for the Uniform, Gaussian and Exponential distributions respectively. The Cost function has been scaled to allow the results for all distributions to be compared.

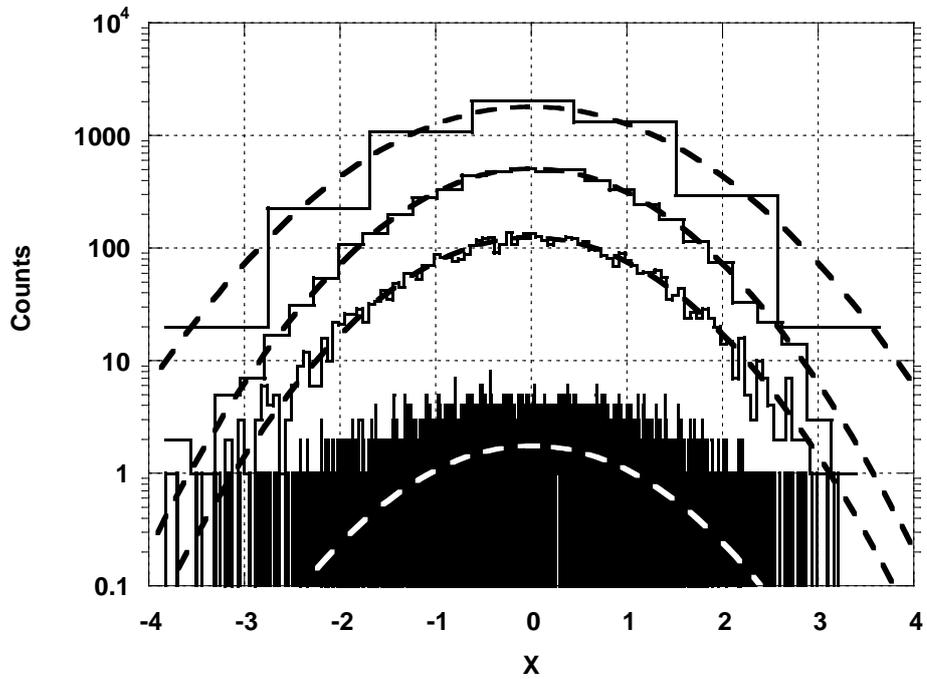

Fig 3.4 Gaussian with mean zero and standard deviation one for different values of M. M is 1, 2, 3, and 6 as the bin size increases and the number of peak counts increase.

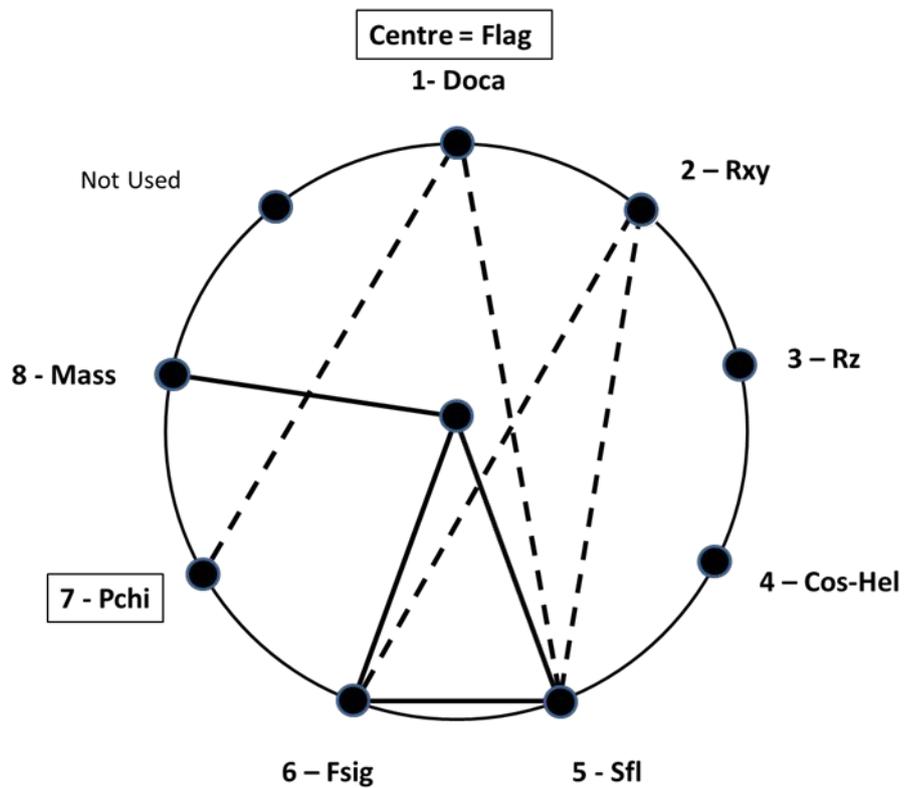

Fig 4.1 Variable Interaction Diagram for the particle physics Monte Carlo dataset described in the text. See text for an explanation.

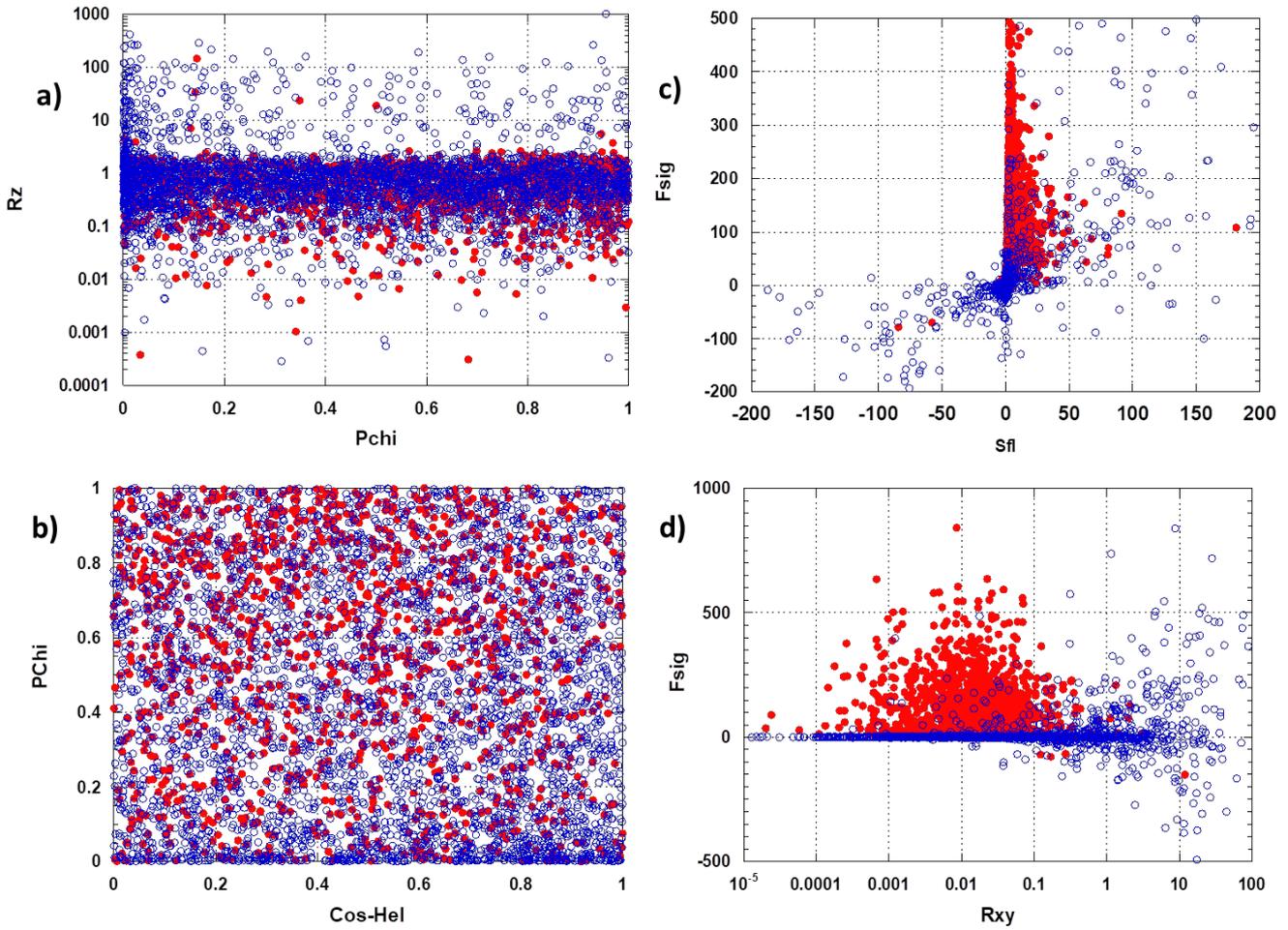

Fig. 4.2 Scatter plots for a) Pchi v Rz, b) Cos-Hel v PChi, c) Sfl v Fsig (c) and d) Rxy v Fsig. Filled points are signal events. Open circles are background events. See text for a discussion. SI = 0, CDI(B,S) = 0.4 bits and CDI(S,B) = 0.2 bits for a). SI, CDI(B,S) and CDI(S,B) are all 0.0 bits for b). SI = 0.25, CDI(B,S) = 4.45 bits and CDI(S,B) = 4.17 bits for c). SI = 0.04, CDI(B,S) = 4.46 bits and CDI(S,B) = 5.1 bits for d).